\documentclass[runningheads]{llncs}
\usepackage[T1]{fontenc}
\usepackage{graphicx}
\usepackage{hyperref}
\usepackage{color}

\usepackage{amsmath,amsfonts}
\usepackage{algpseudocode}
\usepackage{algorithm}
\usepackage[acronym]{glossaries}
\usepackage{cite}
\usepackage[bottom]{footmisc} 
\usepackage{booktabs}
\usepackage{array} 
\usepackage{amssymb}
\usepackage[font=small, labelfont=bf]{caption}
\usepackage{lmodern}
\usepackage{microtype}
\usepackage{pgf}
\usepackage{xcolor}
\usepackage{tikz}
\usepackage[final]{changes}
\definechangesauthor[name={Chung}, color=red]{C}
\usetikzlibrary{arrows.meta,calc,positioning}

\graphicspath{{figures/tohoku-matrix/}{figures/tohoku-domain/}{figures/tohoku-uq-samples/}}

\algnewcommand\algorithmicparfor{\textbf{parallel for}}
\algdef{SE}[PARFOR]{ParFor}{EndParFor}[1]
  {\algorithmicparfor\ #1\ \textbf{do}}
  {\algorithmicend\ \algorithmicparfor}
\algnewcommand{\algorithmicgoto}{\textbf{go to}}%
\algnewcommand{\Goto}[1]{\algorithmicgoto~\ref{#1}}%

\newacronym{uq}{UQ}{uncertainty quantification}
\newacronym{hpc}{HPC}{high performance computing}
\newacronym{gp}{GP}{Gaussian process}
\newacronym{http}{HTTP}{Hypertext Transfer Protocol}
\newacronym{umbridge}{UM-Bridge}{UQ and Modelling Bridge}
\newacronym{pde}{PDE}{partial differential equation}
\newacronym{mcmc}{MCMC}{Markov chain Monte Carlo}
\newacronym{mlmcmc}{MLMCMC}{multilevel Markov chain Monte Carlo}
\newacronym{mlda}{MLDA}{multilevel delayed acceptance}
\newacronym{da}{DA}{delayed acceptance}
\newacronym{mc}{MC}{Monte Carlo}
\newacronym{mpi}{MPI}{Message Passing Interface}
\newacronym{mh}{MH}{Metropolis--Hastings}
\newacronym{qoi}{QOI}{quantity of interest}
\newacronym{fwi}{FWI}{full waveform inversion}
\newacronym{dof}{DOF}{degrees of freedom}

\newcommand{\ExaHyPE}{ExaHyPE}
\newcommand{\ADERDG}{ADER-DG}
\newcommand{\FV}{FV}
\newcommand{\Q}{Q}
\newcommand{\F}{F}
\newcommand{\B}{B}
\newcommand{\R}{\mathbb{R}}
\newcommand{\bathy}{b}
\newcommand{\Ns}{N_{\mathrm{s}}}
\newcommand{\PiDGtoFV}{\Pi_{\mathrm{DG}\rightarrow\mathrm{FV}}}
\newcommand{\PiFVtoDG}{\Pi_{\mathrm{FV}\rightarrow\mathrm{DG}}}

\newcommand{\anne}[1]{#1}

\glsdisablehyper
\makeglossaries

\begin{document}
\title{Dynamic Load Balancing for Uncertainty Quantification with Applications in Bayesian Inversion}
\titlerunning{Dynamic Load Balancing for Uncertainty Quantification}
%
\author{Chung Ming Loi\inst{1} 
\and Mario Wille\inst{2}
\and Anne Reinarz\inst{1}}
\authorrunning{C. M. Loi et al.}
%
\institute{Durham University, UK\\
\email{\{chung.m.loi,anne.k.reinarz\}@durham.ac.uk}
\and 
Technical University of Munich, Germany\\
\email{mario.wille@tum.de}}

\maketitle              

\begin{abstract}
\Gls{uq} workflows present a particular scheduling challenge in high performance computing environments, as they typically generate large numbers of heterogeneous model evaluations with loose but non-trivial dependencies between tasks. A static one-size-fits-all approach in traditional schedulers is inadequate to handle heterogeneous tasks optimally. We introduce an improved load balancer in the \gls{umbridge} framework aimed at mitigating these issues; \gls{umbridge} is a language-agnostic interface developed to couple \gls{uq} software with numerical simulation. \replaced[id=C]{This load balancer is first benchmarked against the previous implementation using a synthetic problem. Then, as} {As} a realistic example, we test the load balancer with a Bayesian inverse problem solved via multilevel delayed acceptance sampling. The underlying forward problem is a hierarchy of tsunami simulations enabled through ExaHyPE, whose runtimes span several orders of magnitude and loose dependencies between levels make the workload particularly challenging to schedule. Our results indicate the load balancer distributes the sampling requests effectively with an average \replaced[id=C]{idle time}{node idle time} of close to a millisecond, a thousand times lower than the previous version, without any prior assumptions about the workload \added[id=C]{or application}.

\end{abstract}
%
%

\section{Introduction} \label{sec:intro}
%
%


\Gls{hpc} has enabled researchers to tackle once computationally 
intractable problems, including large-scale Bayesian inverse problems in which 
the posterior distribution over unknown parameters must be recovered from 
indirect observations. These problems arise naturally in geophysical 
applications such as seismology and tsunami modelling, where the forward model 
is an expensive numerical simulation of a \gls{pde} \cite{10.1145/3126908.3126948, 10.3389/feart.2020.591549, 10.1145/3712285.3771787}.
Alongside advancements in the hardware landscape, scheduling and resource management are another factor in utilising \gls{hpc} efficiently. Schedulers like SLURM, PBS, and LoadLeveler were designed for traditional \gls{hpc} workloads, where jobs are long-running and homogeneous. However, production clusters at Lawrence Livermore National Laboratory \cite{flux, doi:10.1177/10943420211051039} reported that 48.1\% of jobs belong to batches of at least 100 identical jobs submitted in a short timeframe, signalling the shift in \gls{hpc} workloads. \Acrfull{uq} workflows are a particularly demanding
example of this shift, as they require large numbers of model evaluations with heterogeneous runtimes and, in many cases, non-trivial dependencies between tasks that complicate scheduling further.

Classical inverse problems deterministically apply the inverse map to find the best fit solution that describes the observation. However, they are notoriously ill-posed in the Hadamard sense: a solution may not exist, may not be unique, or may not depend continuously on the data \cite{bay_inverse}. Tikhonov regularisation \cite{tikhonov1963} and related variational 
approaches can restore well-posedness, but yield only a point estimate with no quantification of uncertainty. The Bayesian formulation \cite{bay_inverse, doi:10.1137/23M1556435} addresses both issues by treating the unknown parameters as a random variable, and conditioning it on the prior knowledge and observation data. It regularises 
the problem while recovering a full posterior distribution. 

A common approach to Bayesian inversion is \gls{mcmc}, which constructs a Markov chain whose stationary distribution is the target posterior \cite{bay_inverse}.
A large class of \gls{mcmc} algorithms rely on sampling-based proposals, e.g. random walk Metropolis--Hastings \cite{10.1093/biomet/57.1.97}, the No-U-Turn Sampler (NUTS) \cite{hoffman2014no}, and Hamiltonian Monte Carlo (HMC) \cite{duane_hybrid_1987}, each requiring tens of thousands of forward model evaluations to converge, and some additionally require derivative information such as gradients or Hessians of the forward model. However, regardless of the proposal, \gls{mcmc} faces a severe computational bottleneck in 
most applications: samples are proposed serially, each requiring one forward 
solve of the underlying numerical model, and Monte Carlo convergence is slow \cite{giles2015multilevel, giles2008multilevel}. 
Multilevel methods have emerged to address this bottleneck by exploiting a hierarchy of model approximations with increasing accuracy and cost \cite{doi:10.1137/19M126966X}. In particular, we employ the \gls{mlda} algorithm \cite{lykkegaard} that generalises the \gls{da} algorithm \cite{Christen01122005} by filtering proposed samples through a sequence of coarse models before committing to an expensive fine-level evaluation. While \gls{mlda} substantially improves sampling efficiency, the resulting workload is heterogeneous across levels and carries loose but non-trivial dependencies between proposals at adjacent levels, 
making it challenging to schedule efficiently on \gls{hpc} systems.


A growing number of tools have emerged to remedy the challenges in \gls{hpc} scheduling. HyperQueue \cite{BERANEK2024101814} is a plugin scheduler on top of the native scheduler,\deleted[id=C]{. It is specifically designed} simplifying execution of large workflows on \gls{hpc} systems by coordinating a network of allocated nodes that execute submitted tasks. Self-contained \gls{uq} software ecosystems like the VECMA project \cite{doi:10.1098/rsta.2020.0221} address the scheduling issues by interfacing with tools like QCG-PilotJob \cite{10.1007/978-3-030-77977-1_39} and Dask \cite{dask}, both of which are Python libraries that provide an API to the native scheduler. However, these tools either require intrusive changes to the application code, assume a priori knowledge of task runtimes, or are tightly coupled to a specific programming language or \gls{uq} ecosystem, thus limiting their general applicability.

In this work, we focus on workloads enabled by the \acrfull{umbridge} framework \cite{SEELINGER2024113542, joss}. It is a language-agnostic 
interface designed to link \acrshort{uq} techniques and numerical simulations 
through the \gls{http} client and server architecture, abstracting the 
forward model as a map $\mathbf{\mathcal{F}}:\mathbb{R}^n \rightarrow 
\mathbb{R}^m$ \added[id=C]{such that model evaluations are requests from the client to the server. So, a deep understanding of both the \acrshort{uq} and simulation codebases is no longer necessary to conduct \acrshort{uq} workflows because they can be performed without intrusive modification to the underlying simulation code.} Where available, the interface also supports the exchange of 
derivative information, including Jacobians, gradients, and Hessians of 
$\mathbf{\mathcal{F}}$, thus enabling gradient-based \acrshort{uq} algorithms such 
as HMC and NUTS to be coupled with simulation codes in the same non-intrusive 
way. By separating concerns between the \acrshort{uq} and simulation 
components, \acrshort{umbridge} allows non-experts to scale up their 
applications without modifying either the \acrshort{uq} or simulation code, 
and lets either component be swapped independently. We
introduce an improved load balancer in \acrshort{umbridge} for scaling up 
\acrshort{uq} workloads in SLURM-enabled \acrshort{hpc} clusters that does not 
require system-level changes and makes no prior assumptions about task runtimes 
or dependencies.


As a realistic example, we solve a Bayesian inverse problem following the example in \cite{SCpaper}. Bayesian inversion for tsunami source parameters is an active research area, with recent work demonstrating the feasibility of large-scale inference \cite{HENNEKING2026114682,tsunamiuq1,tsunamiuq2}. We use the ExaHyPE engine to simulate the 2011 T\=ohoku tsunami, caused by an earthquake in the Japan trench. \anne{The forward model is a shallow-water tsunami simulation of the depth-averaged shallow-water equations implemented in ExaHyPE \cite{peano,Reinarz2020ExaHyPE}.} Our inverse problem is to recover the location of the earthquake from the tsunami produced by ExaHyPE and real ocean buoy data. Inference is performed using \acrshort{mlda} with a three-level model hierarchy: the two finest levels are ExaHyPE shallow-water simulations of differing spatial resolutions, following the approach of \cite{SCpaper}, while the coarsest level replaces a further coarse \acrshort{pde} solve with a \gls{gp} surrogate \cite{rasmussen2006gaussian} trained on fine-level simulation output, reducing the cost of the most frequently evaluated level by several orders of magnitude. The three-level model hierarchy, whose runtimes span several orders of magnitude, combined with the loose dependencies between levels inherent to \acrshort{mlda}, provides a demanding test of the load balancer. \added[id=C]{Additionally, we run a synthetic \acrshort{mlda} problem, with similar model hierarchies, to compare the current work against the load balancer in \cite{11018268}}.

The paper is structured as follows. The load balancer design and implementation are presented in Section~\ref{sec:umbridge}. Section~\ref{sec:exahype} introduces ExaHyPE as a framework for simulating first-order hyperbolic \acrshortpl{pde} and describes 
the T\=ohoku tsunami configuration. Section~\ref{sec:inverse problem} formulates the Bayesian inverse problem, and Section~\ref{sec:mlda} describes the \acrshort{mlda} algorithm. Finally, Section~\ref{sec:results} presents numerical results demonstrating both the efficacy of the load balancer and the recovered posterior distribution for the tsunami source location, before we conclude in Section~\ref{sec:conclusion}.


\section{SLURM-Integrated Load Balancing in UM-Bridge} \label{sec:umbridge}


We introduce an improved \acrshort{umbridge} load balancer that directly 
addresses the scheduling challenges of \acrshort{mlda} workloads identified in 
Section~\ref{sec:intro}, in particular the elimination of per-request 
initialisation overhead that made existing implementations \cite{11018268} 
unsuitable for heterogeneous task durations.

\subsection{Architecture Overview}
\acrshort{umbridge} \cite{joss, SEELINGER2024113542} is a 
language-agnostic interface that connects \acrshort{uq} algorithms with 
numerical models non-intrusively via an \acrshort{http} client-server 
architecture. The forward model is abstracted as a map 
$\mathcal{F}:\mathbb{R}^n \rightarrow \mathbb{R}^m$, evaluated at points 
$\{\boldsymbol{\theta}_i\}_{i=0}^N$ that may be predetermined or determined 
adaptively, as in Bayesian inversion. \added[id=C]{Most importantly, the \acrshort{uq} algorithm is isolated in the client, and the numerical model in the server, so model evaluations are requests from the client to the server, without integrating both components into the same codebase.} \deleted[id=C]{Where available, \acrshort{umbridge} also supports the exchange of derivative information—Jacobians, gradients, and Hessians of $\mathcal{F}$—enabling gradient-based \acrshort{mcmc} 
algorithms such as HMC and NUTS to be coupled with simulation codes in the same non-intrusive way.} By treating \acrshort{uq} algorithms and numerical solvers as entirely separate applications, \acrshort{umbridge} allows components written in different programming languages to be coupled and swapped out without modification to either side, lowering the barrier to non-experts scaling up their applications. The load balancer 
routes evaluation requests $\{\mathcal{F}(\boldsymbol{\theta}_i)\}_{i=0}^N$ 
across available servers; since all data dependencies are managed by the 
\acrshort{uq} client, the load balancer requires no knowledge of the 
\acrshort{uq} algorithm, model structure, or expected runtimes, making it 
broadly applicable beyond \acrshort{uq} to any setting where \replaced[id=C]{the client-server architecture is appropriate.}{client and server components are separable.}



\subsection{SLURM-Integrated Load Balancer} \label{subsec:LB}

\acrshort{uq} workflows based on sampling create a distinctive scheduling 
problem: they submit large numbers of heterogeneous tasks in a short timeframe 
\cite{Farcas2022AGF}, with often unpredictable runtimes spanning
several orders of magnitude. Traditional SLURM-based approaches struggle here
for two reasons. First, each job submission incurs launch overhead that 
can dominate the runtime of short-lived tasks. Second, users must specify a 
time limit per job, which is typically set conservatively to the worst-case 
runtime, leading to systematic underutilisation \added[id=C]{due to the backfilling algorithm} when most tasks finish much
earlier. For \acrshort{mlda} workloads, these problems are compounded by the 
sequential dependencies between levels: a fine-level evaluation cannot 
proceed until the coarse filter has accepted a proposal, so idle time at one 
level directly delays progress at the next.

The \acrshort{umbridge} load balancer is an abstraction layer between the parallel \acrshort{uq} client and model servers running on \acrshort{hpc} compute nodes. It forwards client requests to available servers like a network proxy. The two existing \acrshort{umbridge} load balancers \cite{11018268} 
address job submission overhead partially, but both initialise a new 
\acrshort{umbridge} server per client request. This per-request initialisation 
cost is acceptable for long-running simulations, but becomes the dominant 
overhead for short tasks such as the \acrshort{gp} surrogate evaluations in our 
\acrshort{mlda} hierarchy, rendering them unsuitable for
heterogeneous task durations.


\begin{figure}[!htb]
    \centering
    \includegraphics[width=0.95\textwidth]{figures/umbridge/CP113_pure-slurm.pdf}
    \caption{Load balancer with multiple \acrshort{umbridge} model instances and a parallel client.}
    \label{fig:umbridge-setup}
\end{figure}

Our new implementation eliminates this bottleneck by using the SLURM job 
array functionality to request one bulk allocation at initialisation and 
persistently manage a pool of \acrshort{umbridge} model servers for the entire 
duration of the run, as shown in Fig.~\ref{fig:umbridge-setup}. Client 
requests are distributed across the pool in a first-come, first-served (FCFS), 
round-robin fashion, with no assumption made about task duration or 
inter-task dependencies. The persistent server pool means that even 
millisecond-duration \acrshort{gp} evaluations are dispatched and returned with 
overhead limited to \acrshort{http} communication latency, while the same 
infrastructure simultaneously handles hour-long fine-grid \acrshort{pde} 
solves.



Concurrent requests are handled safely using a \texttt{mutex} lock and 
\texttt{std::queue} to preserve arrival order and prevent race conditions. A 
\texttt{std::condition\_variable} allows threads to sleep and wake without 
polling: whenever a server completes a request and is marked free, a 
\texttt{notify\_all()} call wakes all queued threads and the next request is 
dispatched immediately. This design keeps the average inter-request idle 
time at the scale of \acrshort{http} communication overhead, as confirmed by 
the experimental results in Section~\ref{sec:results}. Algorithm~\ref{alg:LB} 
describes the dispatch logic. The implementation requires no admin 
privileges and does not modify the native scheduler, making it deployable on 
any SLURM-enabled cluster without system-level changes. It also works with existing \acrshort{umbridge} setups unmodified.
\begin{algorithm}[!htb]
\caption{Simplified algorithm describing the UM-Bridge load balancer.}
\begin{algorithmic}[1]

\ParFor{$j = 0$ to $N-1$} \Comment{Concurrent incoming requests}
    \State \texttt{mutex.lock()}
    \State \texttt{queue.push(request[$j$])}

    \If{free server exists} \Comment{Point of entry after wakeup} \label{checkpoint} 
        \State \texttt{server} $\gets$ \texttt{getFreeServer()}
        \State \texttt{request[$j$]} $\gets$ \texttt{queue.pop()}
        \State \texttt{server.markBusy()}
        \State \texttt{mutex.unlock()}
        \State \Return \texttt{server(request[$j$])} \Comment{Blocking call; reset busyness once done}
    \Else
        \State \texttt{condition\_variable.wait(mutex)}
            \Comment{Sleep current thread}
        \State \Goto{checkpoint} \Comment{Triggered whenever a server is marked as free}
    \EndIf
\EndParFor
\end{algorithmic}
\label{alg:LB}
\end{algorithm}

\added[id=C]{The main bottleneck of this FCFS load balancer design stems from the \texttt{mutex}, which forces the server dispatch to be serial. Updating the queue takes constant time, but selecting a free server requires looping through the list of UM-Bridge servers. So the overhead from the load balancer scales linearly with the number of available servers in the worst case due to \texttt{mutex} contention. For MCMC sampling, this limitation is less problematic because the number of parallel chains is usually fixed to the number of servers since only one sample per chain can be computed at one time. Additionally, longer chains are generally preferred over an abundance of shorter ones to achieve a comparable sample count, hence lowering the risk of saturating the workflow with \texttt{mutex} overheads.}


\section{ExaHyPE: Exascale Hyperbolic PDE Engine} \label{sec:exahype}
In this work we use ExaHyPE to simulate the 2011 T\=ohoku tsunami. We briefly describe the numerical scheme and the tsunami application below.

\subsection{ADER-DG and A Posteriori Finite-Volume Subcell Limiting}

\ExaHyPE{}\footnote{Available as Docker image at \url{https://hub.docker.com/r/peanoframework/exahype2}} is a simulation framework for first-order hyperbolic balance laws
with conservative fluxes, non-conservative products, and source terms,
\begin{equation}
  \partial_t \Q
  + \sum_{d=1}^{D}\partial_{x_d}\F_d(\Q)
  + \sum_{d=1}^{D}\B_d(\Q)\,\partial_{x_d}\Q
  = S(\Q,x,t),
  \qquad
  \Q\in\R^m ,
  \label{eq:balance-law}
\end{equation}
using the engine design described in \cite{Reinarz2020ExaHyPE}.
The regular discretisation is an \ADERDG{} method, i.e.\ a discontinuous Galerkin (DG) scheme
with arbitrary high-order derivative (ADER) time integration. On every cell $K$, the state
is a tensor-product polynomial $Q_h|_K\in V_h^p(K)$. A local space-time
predictor evolves this polynomial inside the cell, and a corrector couples
neighbouring cells through face fluxes and non-conservative interface terms.
This yields a high-order one-step method with low numerical diffusion in smooth
solution regions \cite{Dumbser2008Unified,Dumbser2018ExaHyPE}.

High-order polynomials are not monotone near shocks, discontinuities, steep
gradients, or under-resolved features. The method thus uses a posteriori limiting: an
\ADERDG{} step proposes $Q_h^{*,n+1}$, which is accepted only
if it satisfies physical admissibility and a relaxed discrete maximum principle
\cite{DumbserLoubere2016}.
\deleted[id=C]{For an observable $w(\Q)$, the maximum principle compares the candidate with
extrema from the previous accepted state on $K$ and its face-neighbourhood
$\mathcal N(K)$,
\begin{equation}
  m_{\mathcal N}^n - \delta
  \le
  w(Q_h^{*,n+1}(x))
  \le
  M_{\mathcal N}^n + \delta,
  \qquad
  \delta =
  \max\!\left(\epsilon,\alpha(M_{\mathcal N}^n-m_{\mathcal N}^n)\right).
  \label{eq:dmp}
\end{equation}}
The admissibility test enforces problem-specific invariant-domain constraints,
such as finite states and positivity of selected components. Accepted
cells stay on the \ADERDG{} layer; rejected cells are recomputed on a robust
subcell finite-volume (\FV{}) layer.

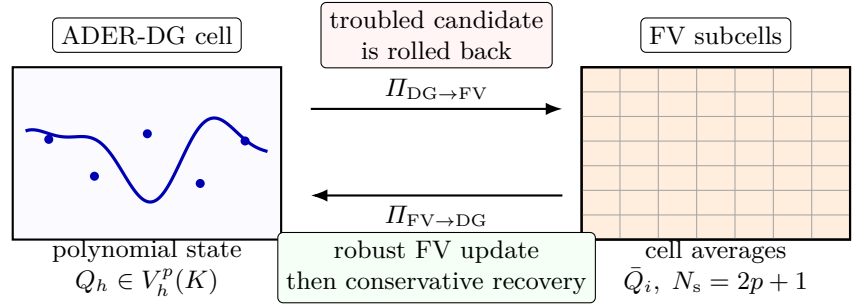
\begin{figure}[bht]
  \centering
  \begin{tikzpicture}[
      scale=1.06,
      transform shape,
      x=1cm,y=1cm,
      every node/.style={font=\small},
      state/.style={draw,rounded corners=2pt,fill=white,inner sep=3pt},
      arrow/.style={-{Latex[length=2.2mm]},thick},
      backarrow/.style={{Latex[length=2.2mm]}-,thick}
    ]
    \begin{scope}[shift={(0,0)}]
      \draw[thick,fill=blue!2] (0,0) rectangle (3.35,2.15);
      \draw[very thick,blue!70!black,domain=0.16:3.18,samples=80,smooth]
        plot (\x,{1.05 + 0.42*sin(165*\x) + 0.18*cos(305*\x)});
      \foreach \x/\y in {0.45/1.25,1.02/0.79,1.68/1.32,2.34/0.70,2.90/1.23}
        \fill[blue!70!black] (\x,\y) circle (1.6pt);
      \node[state] at (1.675,2.55) {\ADERDG{} cell};
      \node[align=center] at (1.675,-0.35) {polynomial state\\$Q_h\in V_h^p(K)$};
    \end{scope}

    \begin{scope}[shift={(7.1,0)}]
      \draw[thick,fill=orange!13] (0,0) rectangle (3.35,2.15);
      \foreach \i in {1,...,6} {
        \draw[gray!70] ({3.35*\i/7},0) -- ({3.35*\i/7},2.15);
        \draw[gray!70] (0,{2.15*\i/7}) -- (3.35,{2.15*\i/7});
      }
      \draw[thick] (0,0) rectangle (3.35,2.15);
      \node[state] at (1.675,2.55) {\FV{} subcells};
      \node[align=center] at (1.675,-0.35) {cell averages\\$\bar Q_i,\; \Ns=2p+1$};
    \end{scope}

    \draw[arrow] (3.72,1.63) -- node[above] {$\PiDGtoFV$} (6.86,1.63);
    \draw[backarrow] (3.72,0.55) -- node[below] {$\PiFVtoDG$} (6.86,0.55);

    \node[state,align=center,fill=red!4] at (5.29,2.55)
      {troubled candidate\\is rolled back};
    \node[state,align=center,fill=green!4] at (5.29,-0.35)
      {robust FV update\\then conservative recovery};
  \end{tikzpicture}
  \vspace{-0.4em}
  \caption{A posteriori subcell limiter. A rejected
  \ADERDG{} candidate is replaced by an \FV{} recomputation on subcell
  averages and projected back to the DG polynomial space.}
  \label{fig:dg-fv-projection}
\end{figure}

For a degree-$p$ \ADERDG{} polynomial, the limiter uses
$\Ns = 2p+1$
subcells per coordinate direction. This samples all polynomial modes and keeps
the \FV{} and \ADERDG{} time-step restrictions compatible
\cite{DumbserLoubere2016}. If
$\omega_i\subset K$ is an \FV{} subcell, the rollback state is projected from
the accepted DG polynomial to subcell averages,
\begin{equation}
  \bar Q_i^n
  =
  \frac{1}{|\omega_i|}
  \int_{\omega_i} Q_h^n(x)\,dx,
  \qquad
  \bar Q^n = \PiDGtoFV Q_h^n .
  \label{eq:dg-to-fv}
\end{equation}
The \FV{} method recomputes the same time interval on these subcells. The
accepted subcell solution is then projected back to $V_h^p(K)$ by a constrained
conservative least-squares reconstruction \cite{DumbserLoubere2016}.
The $\Ns=2p+1$ \FV{} subcell values per coordinate direction are
used to recover only $p+1$ DG coefficients. The
\FV{}-to-DG reconstruction is therefore overdetermined, so a constrained
least-squares reconstruction is used, with conservation enforced through a
Lagrange multiplier:
\begin{equation}
  Q_h^{n+1}
  =
  \PiFVtoDG\bar Q^{n+1},
  \qquad
  \int_K Q_h^{n+1}\,dx
  =
  \sum_i |\omega_i|\bar Q_i^{n+1}.
  \label{eq:fv-to-dg}
\end{equation}
The resulting method is locally hybrid: high-order \ADERDG{} is retained where 
the solution is smooth, while the \FV{} subcell update supplies robustness near shocks, discontinuities, and other troubled regions.
The limiter acts only where the a posteriori admissibility
check fails: the cost of the \FV{} recomputation is therefore
confined to a small fraction of cells per time step, preserving the
efficiency of the high-order scheme globally.
Figure~\ref{fig:dg-fv-projection} summarises the rollback, subcell update, and
conservative recovery between both representations.

\vspace{-0.3em}
\vspace{-0.4em}

\subsection{2011 T\=ohoku Tsunami}
The tsunami is modelled with the depth-averaged shallow-water equations over a
variable bathymetry $\bathy(x)$. We use the basic shallow-water equations with bathymetry
source terms and neglect bottom friction as well as non-hydrostatic
corrections\replaced[id=C]{. While the mathematical details are unchanged from \cite{SCpaper}, the implementation has since been revamped and is presented for the first time in this work.}{\mbox{\cite{SCpaper}}.} With water depth $h$, horizontal velocity $(u,v)$, momenta $(hu,hv)$, gravity $g$, and free surface $\eta=h+\bathy$, the system is written
as
\begin{equation}
  \partial_t
  \begin{pmatrix}
    h\\ hu\\ hv\\ \bathy
  \end{pmatrix}
  +
  \nabla\!\cdot
  \begin{pmatrix}
    hu  & hv\\
    hu^2 & huv\\
    huv & hv^2\\
    0 & 0
  \end{pmatrix}
  +
  \begin{pmatrix}
    0\\
    gh\,\partial_x(\bathy+h)\\
    gh\,\partial_y(\bathy+h)\\
    0
  \end{pmatrix}
  =0 .
  \label{eq:shallow-water}
\end{equation}

The bathymetry is carried as a time-independent state component, with the 
pressure gradient and bed slope represented together through 
$gh\nabla(\bathy+h)$, exposing the lake-at-rest balance $(u,v)=0$, 
$\eta=\mathrm{const.}$ that the numerical method must preserve while allowing 
large bathymetry jumps, dry land, and inundation. The T\=ohoku configuration 
uses the domain $[-499,1299]\times[-949,849]\,\mathrm{km}$ around Japan, with 
the ocean initially at rest and the earthquake displacement introduced as a 
filtered change of bed elevation. The plotted quantity is the sea-surface-height 
anomaly (SSHA), i.e.\ the deviation of $\eta$ from the still-water state.

Although $\bathy$ is constant in time, it is carried as an unknown
in both the \ADERDG{} and \FV{} layers so limiter projections act on the
full state $(h,hu,hv,\bathy)$. This is required for consistency: since 
$\eta=h+\bathy$ and the bed-slope source depend on $h$ and $\bathy$ together, 
treating them independently across layers would destroy the lake-at-rest 
balance, introduce artificial SSHA, and trigger spurious limiting.

The \FV{} layer uses an augmented-state Riemann solver \cite{George2008}
that decomposes a jump augmented with the momentum flux and bathymetry step,
balancing the bed-slope term inside the interface solve, not as a
separate correction. This preserves well-balancedness over rough bathymetry, 
enforces non-negative water depth, and handles wet/dry interfaces as inundation 
problems, with one-sided draining capped so no \FV{} update removes more
water than is available locally.

\begin{figure}[!bht]
  \resizebox{\textwidth}{!}{%
    \input{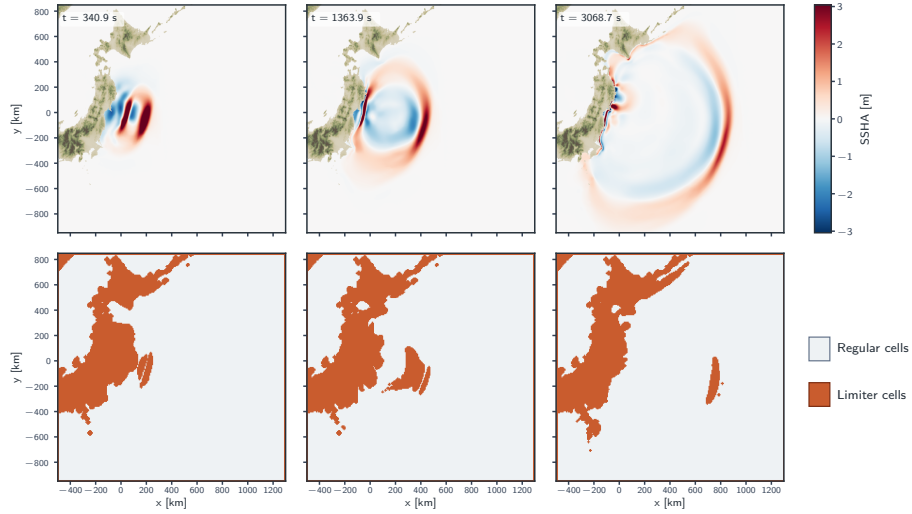}%
  }
  \caption{T\=ohoku tsunami snapshots at
  $t=340.9\,\mathrm{s}$, $1363.9\,\mathrm{s}$, and
  $3068.7\,\mathrm{s}$. Top: SSHA at these times. Bottom: Cells treated by the
  finite-volume limiter for the same times.}
  \label{fig:tohoku-matrix}
\end{figure}

\deleted[id=C]{For this scenario, the limiter is deliberately geometry dominated. A DG cell remains on the \ADERDG{} layer only if the state is finite, the water column 
exceeds the wet/dry threshold, the cell lies in deep water, and the cell-local 
variation of $\eta$ is small relative to the local depth; coastline, wet/dry, 
source-region, and boundary cells are therefore advanced through the \FV{} 
limiter. Boundary cells are not physically troubled but are limited because the 
a posteriori troubledness test is underdetermined there: no exterior DG 
neighbour polynomial exists, so a one-cell \FV{} layer is used to remove 
one-sided reconstruction artefacts while applying the same positivity-preserving 
Riemann treatment used near land and shallow water.}

Figure~\ref{fig:tohoku-matrix} shows three snapshots. The upper row shows
SSHA and the lower row limiter ownership. The open-ocean wave is
advanced by the regular \ADERDG{} scheme; limiter cells are concentrated where
the a posteriori checks are intentionally conservative. Besides the
stationary coastline and boundary layer, a moving band of limiter cells follows
the tsunami wavefront, marking where the evolving free-surface gradients
temporarily require the robust \FV{} update.

\section{The Bayesian Inverse Problem} \label{sec:inverse problem}
A Bayesian inverse problem is the task of reconstructing the posterior 
distribution $\mathcal{P}(\theta \mid y)$ of uncertain parameters
$\theta$ given measurement data $y$ and a forward model
$\mathcal{F}$. By Bayes' theorem,
\begin{equation} \label{eq:bayes}
\mathcal{P}(\theta \mid y) \propto \mathcal{L}(y \mid \theta) \, \anne{\pi_0}(\theta),
\end{equation}
where $\mathcal{L}(y \mid \theta)$ is the likelihood of observing $y$ given $\theta$, and $\anne{\pi_0}(\theta)$ is the prior distribution 
encoding knowledge of $\theta$ before observing data. The 
normalising constant $\mathcal{P}(y)$ is intractable in all but the simplest 
cases and cancels in \acrshort{mcmc} acceptance ratios, so only the 
proportional form \eqref{eq:bayes} is required in practice.


In this work, we are concerned with the 2011 T\=ohoku tsunami, which resulted from an earthquake in the Japan trench. Our inverse problem is recovering the location of the initial displacement from real-world bathymetry\footnote{\url{https://www.gebco.net/}} and ocean buoy\footnote{\url{https://www.ndbc.noaa.gov/}} data. We choose $\pi_0(\theta)$ as a 2D uniform distribution in the translation window $[-200, 200] \times [-200, 200] \, \mathrm{km}$ to reflect zero a priori knowledge about the unknown coordinates with support 
chosen to keep the initial displacement away from the domain boundary and sufficiently far from the probes to avoid unphysical simulation output. Figure~\ref{fig:tohoku-matrix} also serves as a reference for the domain coordinates.

We formulate the inverse problem in accordance with \cite{SCpaper}, where the Gaussian likelihood is chosen to quantify the mismatch between observation and simulation. The mean vector contains the wave height and arrival time of 
the tsunami at the two Deep-ocean Assessment and Reporting of Tsunamis (DART) probes 21418 and 21419, while the diagonal 
covariance matrix encodes both measurement noise in the probes and the 
discrepancy between the numerical model and reality.

\section{Multilevel Delayed Acceptance MCMC} \label{sec:mlda}
\acrshort{mcmc} methods construct a Markov chain on the parameter space such that the chain converges iteratively to the target distribution, allowing posterior quantities of interest to be estimated from its samples. The \acrfull{mh} algorithm \cite{10.1063/1.1699114, 10.1093/biomet/57.1.97} is the canonical method to construct such chains, but its efficiency depends heavily on the choice of proposal distribution. Many alternatives have been proposed to increase sampling efficiency. We now describe the classic \acrshort{da} algorithm proposed by \cite{Christen01122005} and its multilevel extension due to \cite{lykkegaard}.

The original \acrshort{da} algorithm is a two-level method that aims to increase the sampling efficiency. It introduces a step that filters out bad proposals early using a coarse model before the expensive evaluation at the fine level. As such, the fine level has higher accuracy but at the cost of time to solution. Conversely, the coarse level is relatively inaccurate but fast to compute. Common choices for the coarse level include surrogate models developed using a collection of simulation outputs, or simply the fine model with a coarser computational domain.

\deleted[id=C]{Algorithm~\ref{alg:da} describes the \acrshort{da} sampler. }Given the current state $\theta^j$, rather than evaluating each proposal directly with the expensive fine model, a preliminary \acrshort{mh} step targeting the cheap coarse density $\pi_C(\cdot)$ filters out poor 
proposals before they reach the fine level. The surviving state $\psi$ is 
accepted or rejected at the fine level with probability
\[
  \alpha_F(\psi \mid \theta^j) = \min\left(1,\, \frac{\pi_F(\psi)\,\pi_C(\theta^j)}{\pi_F(\theta^j)\,\pi_C(\psi)}\right),
\]
\anne{which corrects for the discrepancy between the coarse and 
fine densities, and preserves the correct stationary distribution.} The key 
computational saving arises because any proposal rejected at the coarse stage 
never triggers a fine-model evaluation, yielding a net reduction in the 
number of expensive forward solves relative to standard \acrshort{mh} that 
grows with the rejection rate of the coarse level.


    
    
    

\begin{figure}[!htb]
    \centering
    \includegraphics[width=0.7\textwidth]{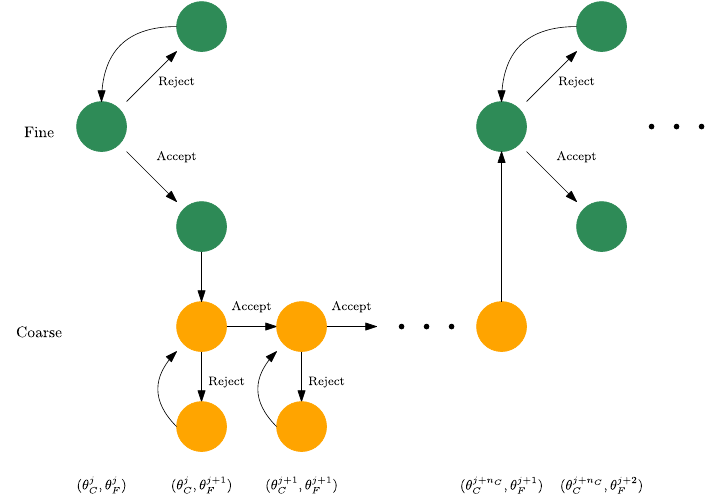}
    \caption{Two-level accept-reject scheme of the \acrshort{mlda} algorithm.}
    \label{fig:mlda tree}
\end{figure}

While \acrshort{da} samples more efficiently than
standard \acrshort{mh}, the acceptance rate at the fine level is 
fundamentally limited by how well the coarse model approximates the fine 
density. Multilevel methods address this by defining a hierarchy of models 
with increasing fidelity, so each level proposes to the next using
a progressively better approximation of the target. A proposed sample passes
through several coarse filters before reaching the expensive fine model,
each filter improving the proposals seen at the next level. This multilevel structure yields variance reduction in the estimators 
at finer levels, as the coarse levels absorb the bulk of the exploration cost 
while passing only well-informed proposals upward.

Figure~\ref{fig:mlda tree} pictorially describes a two-level \acrshort{mlda} chain, which generalises \acrshort{da} by replacing the single coarse step with a randomised subchain 
of length $n_\ell \sim p_\ell(\cdot)$, allowing different state spaces at 
adjacent levels, and generating each subchain recursively via \acrshort{mlda} 
at level $\ell-1$, reverting to \acrshort{mh} at $\ell=0$.

\section{Tsunami Inversion and Load Balancer Performance} \label{sec:results}


\added[id=C]{The tsunami simulations ran on the DINE2 partition of the COSMA cluster. It is equipped with two Intel Xeon Gold 6430 32-core processors and 2\,TB of RAM per node. The synthetic problem ran on the Hamilton8 cluster that has two AMD EPYC 7702 CPUs and 256\,GB of RAM per standard node. Both clusters use HDR200 InfiniBand and are hosted at Durham University. We requested five concurrent \acrshort{umbridge} servers (in a 5-element job array) to host five parallel \acrshort{mlda} chains for both experiments. Each server for the synthetic problem was allocated 1 CPU core and 4\,GB RAM, whereas the tsunami simulation received 2 MPI ranks, 16 OpenMP threads per rank, and 32\,GB RAM per server. The nodes were not exclusive and no special priority was requested, reflecting a realistic production environment.} \acrshort{mlda} sampling is facilitated by the \texttt{tinyDA} Python package\footnote{\url{https://github.com/mikkelbue/tinyDA/tree/main}}, showcasing \acrshort{umbridge}'s ability to couple \acrshort{uq} and simulation software written in different programming languages. All code to reproduce these experiments is available on Zenodo\footnote{\url{https://doi.org/10.5281/zenodo.23420609}}.


\subsection{Synthetic MLDA Problem}
\begin{figure}[!htb]
    \centering
    \includegraphics[width=0.49\textwidth]{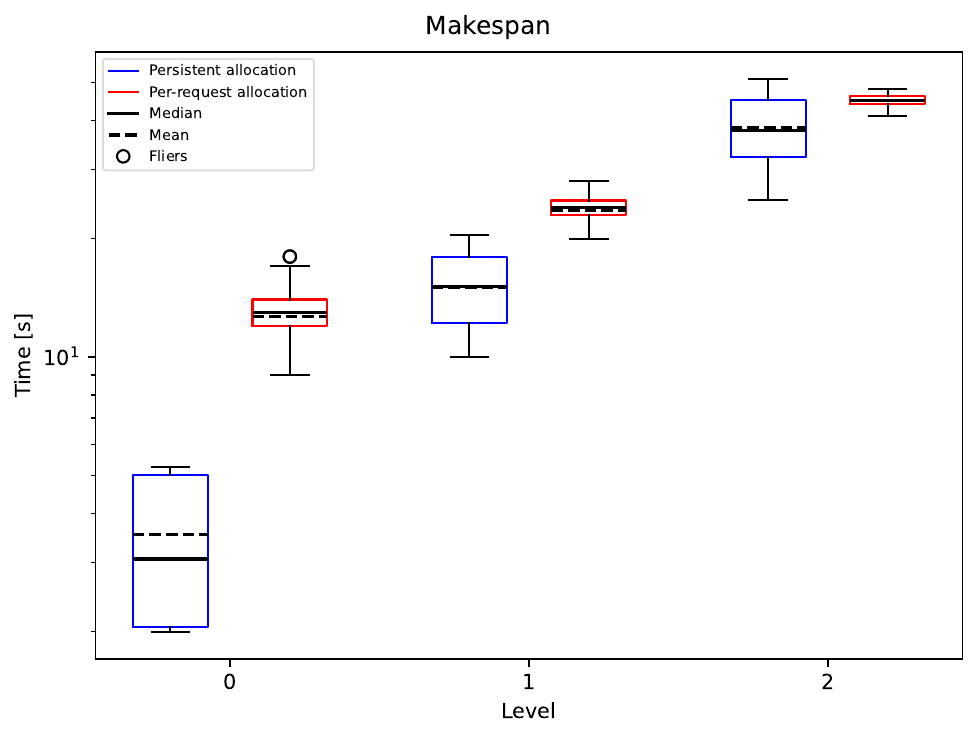}
    \includegraphics[width=0.49\textwidth]{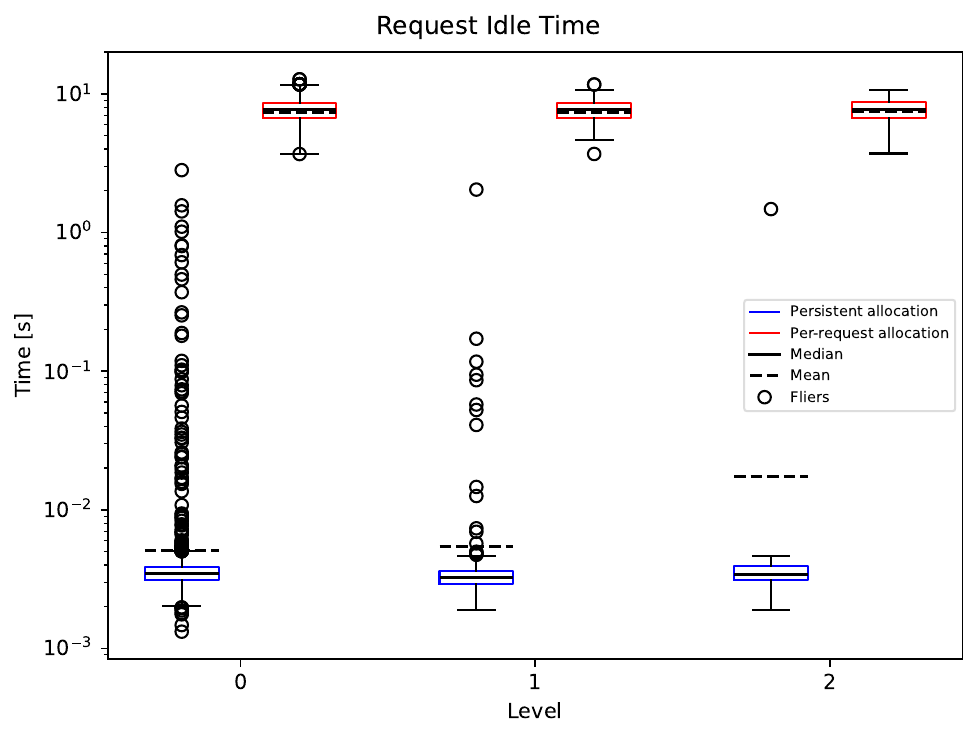}
    \caption{Makespan (left) and idle time (right) of individual requests per MLDA level.
    }
    \label{fig:toy problem timings}
\end{figure}

\added[id=C]{We first present experiment timings in Fig.~\ref{fig:toy problem timings}, comparing two modes of sample evaluation through the predator-prey example in the \texttt{tinyDA} repository. The first mode is the na\"ive SLURM approach (labelled ``per-request allocation'') as discussed in \cite{11018268}, where \texttt{sbatch} is called for every sample. The second mode is the persistent server approach (labelled ``persistent allocation'') as outlined in Section~\ref{subsec:LB}. The timings were obtained by recording the arrival and departure times of individual requests. Since data dependencies are managed by the \texttt{tinyDA} client, and the load balancer greedily assigns parallel requests to available \acrshort{umbridge} servers, the timestamps directly represent time spent inside a server and—by implication—the idle time between requests.}

\added[id=C]{The synthetic problem was injected with artificial compute load to control the CPU time per level, which is reflected in the makespan as CPU time increases with the level. To clarify, makespan measures the total time from submission to completion of a request, including the queue time. We use request idle time as an umbrella term to encapsulate various overheads in the workflow; it is simply the difference between makespan and CPU time. The idle times remain relatively constant across levels because they are not directly affected by the compute cost of a sample. Notably, the persistent server approach has consistently lower idle times—around a thousand times lower—than the previous load balancer, which is also reflected in the makespan. While idle times in Fig.~\ref{fig:toy problem timings} may seem insignificant, they can add up substantially over many samples. In particular, this synthetic workflow had around 10{,}000 combined samples across five chains, leading to a total runtime of 15.5\,h (persistent allocation) and 42.8\,h (per-request allocation) where the discrepancy arises due to the accumulation of overheads.}

\subsection{Inverse Problem: T\=ohoku Tsunami}
The \acrshort{mlda} algorithm was configured with three levels: level~2 is a 
fine-grid ExaHyPE model, level~1 is a coarse-grid ExaHyPE model, and level~0 
is a \acrshort{gp} surrogate trained on 512 Latin hypercube samples drawn from 
the level~1 model. The \acrshort{gp} was implemented in PyTorch with a $5/2$ 
Mat\'ern kernel, zero mean, and automatic relevance determination; 
hyperparameters were optimised by maximising the marginal likelihood on the 
training data. The $\mathcal{O}(n^3)$ cost of kernel matrix operations is 
negligible at $n=512$ relative to the \acrshort{pde} solves at finer levels. 
The \acrshort{mlda} setup is summarised in Table~\ref{tab:setup and mlda 
result}.
\begin{table}[!htb]
    \centering
    \begin{tabular}{c c c c m{1.5cm} m{1.5cm} m{1.5cm} m{1.5cm}}
        \hline
        \multicolumn{1}{b{1cm}}{\centering Level $\ell$} &
        \multicolumn{1}{b{1cm}}{\centering $\bar{t}_\ell$ [s]} &
        \multicolumn{1}{b{1cm}}{\centering $h_\ell$ [m]} &
        \multicolumn{1}{b{1cm}}{\centering DOF} &
        \multicolumn{2}{>{\centering\arraybackslash}m{3cm}}{\centering $\mathbb{E}[\phi_\ell]$} &
        \multicolumn{2}{>{\centering\arraybackslash}m{3cm}}{\centering $\mathbb{V}[\phi_\ell]$} \\
        \hline
        0 & 0.03 & N/A & 512 & -204.45  & -7962.52 & 20329.02  & 43285.31 \\
        1 & 143.03 & 22197.50 & 656100 & -6908.64 & -2767.88 & 13279.11 & 28717.72 \\
        2 & 3071.53 & 7399.18 & 5904900 & -2638.03 & -10607.09 & 8680.50 & 15659.04 \\
        \hline
    \end{tabular}
    \caption{\acrshort{mlda} model hierarchies and inverse problem results. The $h_\ell$ column lists the cell size of the computational domain. The \acrfull{dof} at level 0 is defined as the dimension of the kernel matrix, whereas for the remaining two levels it is the number of \acrshort{dof} in the DG solver. The expectation (in metres) and variance are given for both coordinates of the initial displacement: $x$ and $y$.}
    \label{tab:setup and mlda result}
\end{table}

\begin{figure}[!htb]
    \centering
    \includegraphics[width=0.48\textwidth]{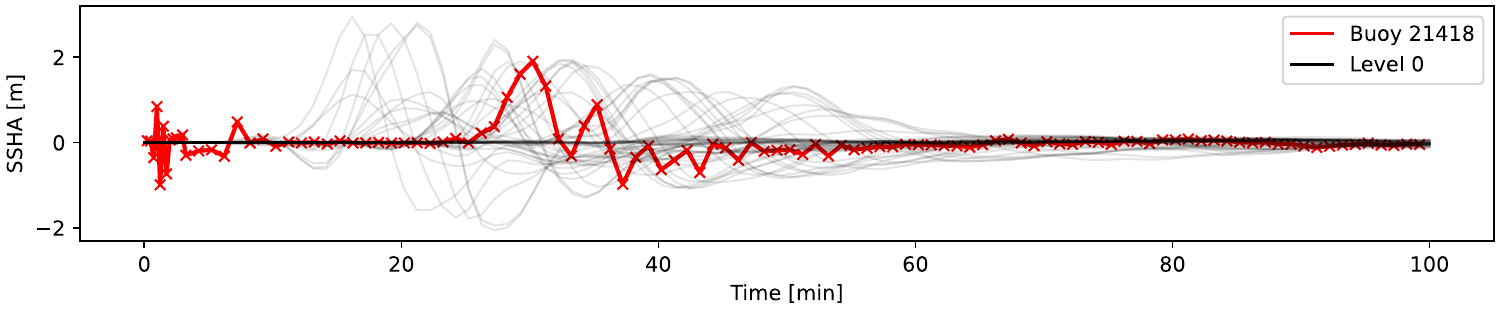}
    \includegraphics[width=0.48\textwidth]{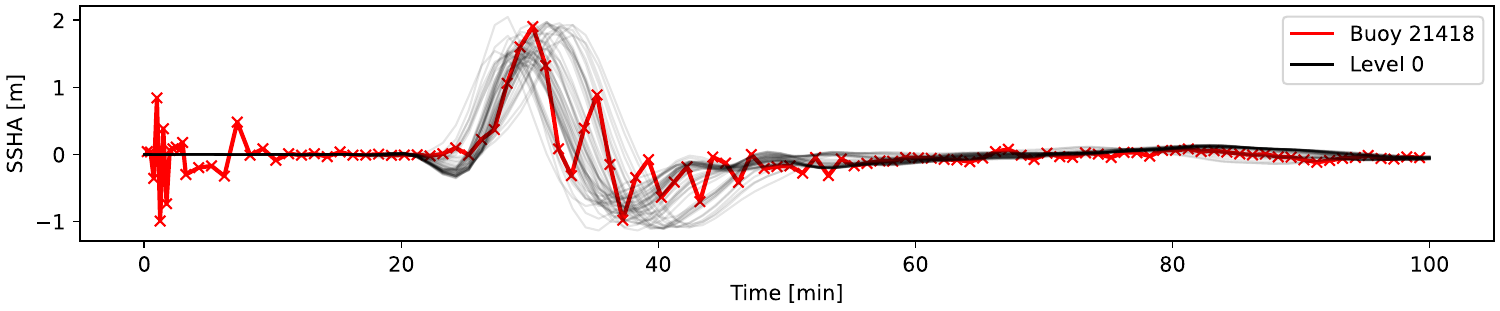}
    \caption{Raw time-series data from probe 21418 and samples at level 0. Left: 50 sample draws from the uniform prior. Right: 50 sample draws from the recovered posterior. 
    }
    \label{fig:probes}
\end{figure}

The runtimes varied by orders of magnitude, averaging
$0.03\,\mathrm{s}$, $143.03\,\mathrm{s}$, and $3071.53\,\mathrm{s}$
per level. The loose dependencies between levels inherent to 
\acrshort{mlda}, where finer samples depend on the acceptance of coarse 
samples, make this workload particularly challenging to schedule on
\acrshort{hpc} systems. In total, the algorithm evaluated $1{,}500{,}005$ 
level~0, $3{,}005$ level~1, and $155$ level~2 samples. Figure~\ref{fig:probes} shows raw data from probe 21418 overlaid with level~0 prior and posterior draws, using a separate \acrshort{gp} trained to reconstruct the full time series.


\begin{figure}[!htb]
  \centering
  \resizebox{0.8\textwidth}{!}{%
    \input{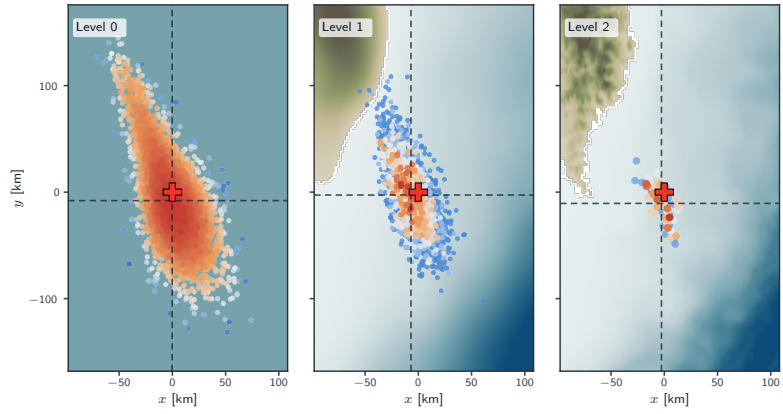}%
  }
  \captionof{figure}{Density of posterior samples from each \acrshort{mlda} level. The dashed lines indicate the sample mean, and the red cross provides a known reference at (0, 0).}
  \label{fig:tohoku-uq-samples}
\end{figure}

Figure~\ref{fig:tohoku-uq-samples} shows the sample densities in each \acrshort{mlda} level. The dashed lines mark the posterior mean $\mathbb{E}[\phi_\ell]$, where $\phi_\ell$ denotes the $(x,y)$ coordinates of the initial displacement sampled at level $\ell$, and the red cross indicates the reference solution at the origin estimated by \cite{10.1093/gji/ggu203}. The primary goal of multilevel methods is to improve acceptance rates through a series of cheap filters, which are corrected progressively with each acceptance at the finer level. \deleted[id=C]{Typically, this effect is reflected by the reduction in variance due to the telescoping sum \eqref{eq:telescoping sum}.} The expectation and variance are listed in Table~\ref{tab:setup and mlda result}. Indeed, we observe variance reduction across levels.\deleted[id=C]{, notably the drop between levels 0 and 1 is larger than expected, since the \acrshort{gp} is conditioned on level 1 samples.}


\subsection{UM-Bridge Load Balancer}

\begin{figure}[!htb]
    \centering
    \includegraphics[width=0.8\textwidth]{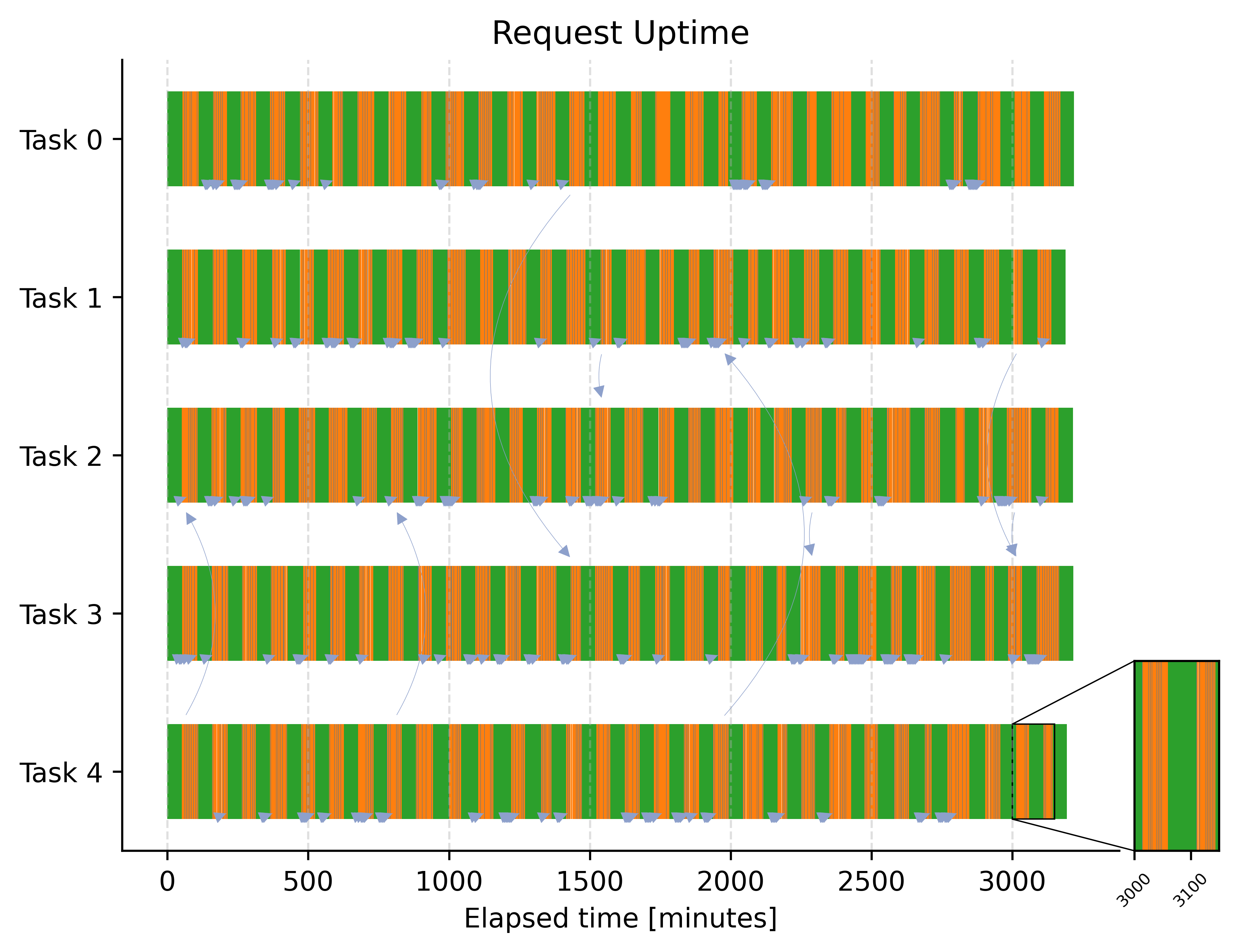}
    \caption{The execution of client requests in the 5-element job array is indicated by the y-axis labels. The x-axis is the elapsed time. The colour of the bars indicates different model fidelities: green is level 2, orange is level 1, and blue is level 0. Arrows indicate the request dependencies within one \acrshort{mlda} chain; the numbers are reduced to avoid clutter.}
    \label{fig:uptime 10956759}
\end{figure}


Figure~\ref{fig:uptime 10956759} shows the uptime of each \acrshort{umbridge} 
server as a bar chart, with colours indicating model fidelity: green for 
level~2, orange for level~1, and blue for level~0. The three-level 
hierarchy presents a demanding scheduling problem: the level~0 
\acrshort{gp} evaluations are five orders of magnitude faster than the level~2 
ExaHyPE runs, yet carry sequential dependencies within each 
\acrshort{mlda} chain; finer level evaluations cannot proceed until the 
coarse filter has accepted a proposal. The level~0 bars are not individually 
visible at this scale due to their negligible runtime of $0.03\,\mathrm{s}$, 
but their dependencies are indicated by the arrows, which trace the request 
ordering within a single \acrshort{mlda} chain. The bars are densely packed 
with little whitespace, indicating that despite task heterogeneity 
the loosely coupled \acrshort{mlda} requests were handled promptly across all 
five parallel chains.

\begin{figure}[!htb]
    \centering
    \includegraphics[width=0.49\textwidth]{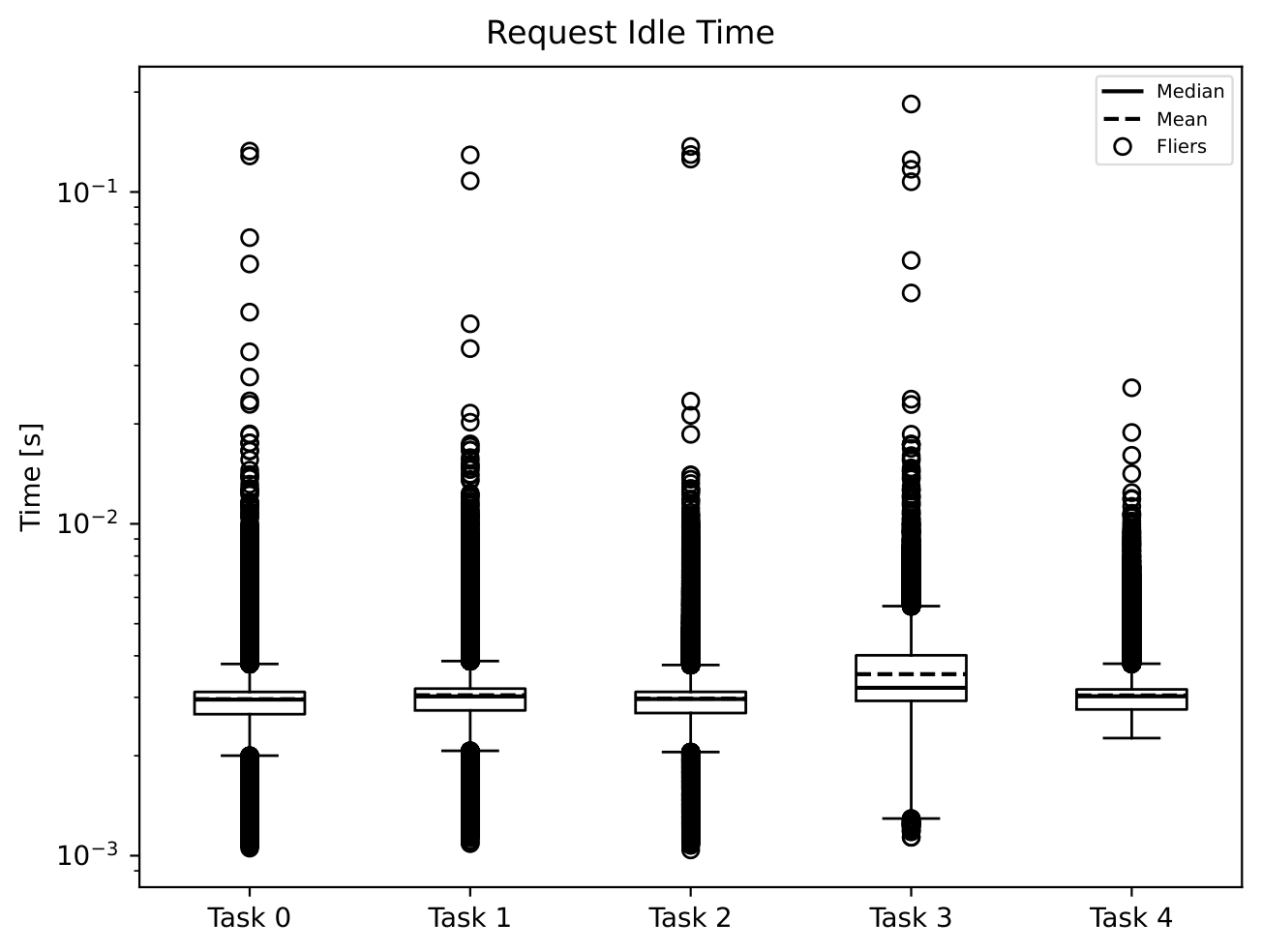}
        \includegraphics[width=0.49\textwidth]{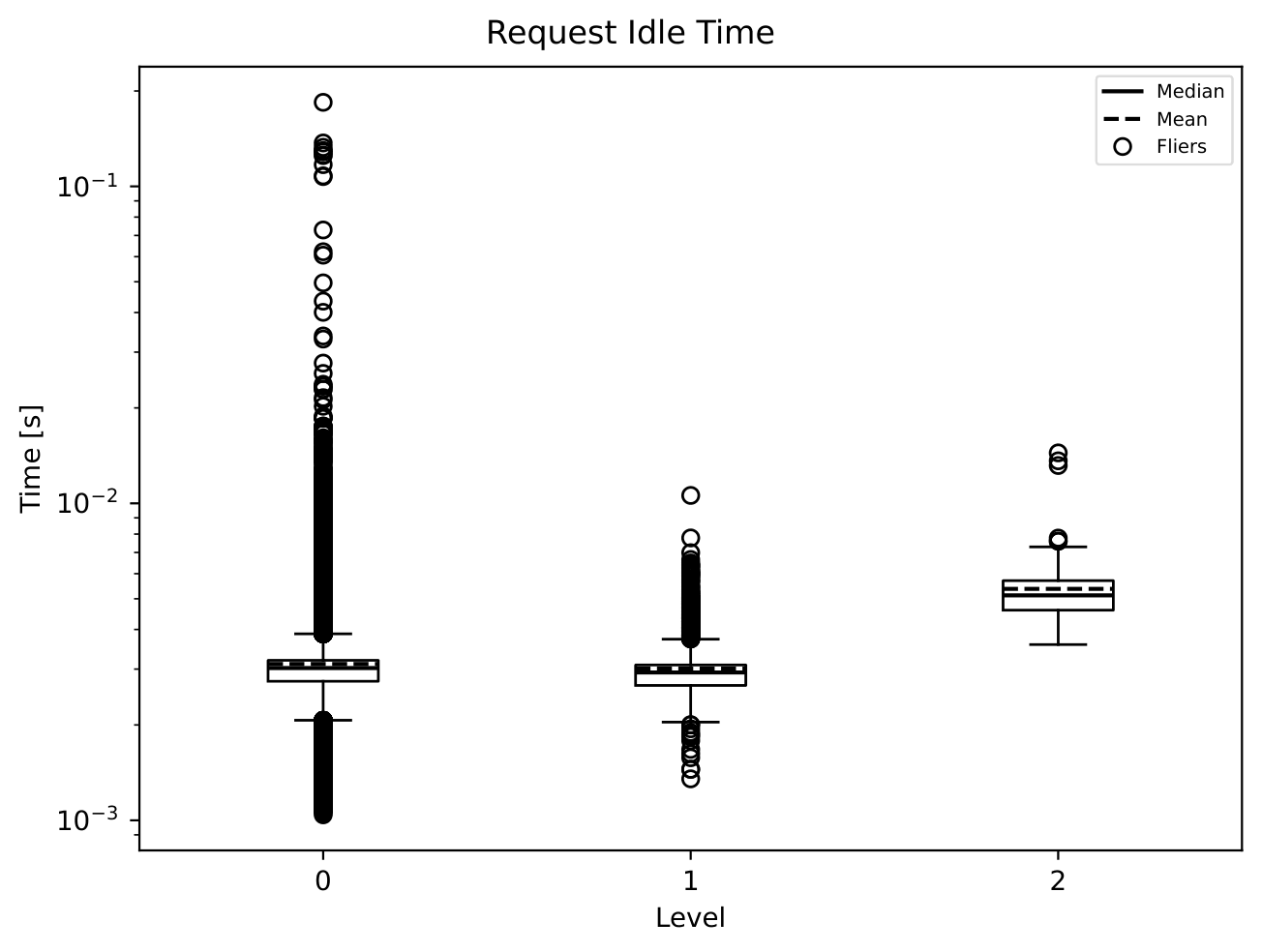}
    \caption{Left: Boxplot of idle time between sampling requests in each persistent server. Right: Breakdown of idle time by level.}
    \label{fig:idle time 10956759}
\end{figure}

The idle times, shown in Fig.~\ref{fig:idle time 10956759}, averaged 
$\mathcal{O}(10^{-3})\,\mathrm{s}$, consistent with the overhead of 
\acrshort{http} communication in \acrshort{umbridge} and substantially lower 
than the per-request server initialisation overhead incurred by the existing 
\acrshort{umbridge} load balancers \cite{11018268}, which made them 
ineffective for short-running tasks such as the level~0 \acrshort{gp} 
evaluations here. \replaced[id=C]{The majority of the outliers originated from level 0, but this is not unusual given that 99.8\% of the overall samples were \acrshort{gp} evaluations. The variation is attributed to occasional network congestion or requests blocked pending the completion of a dependency.}{Outliers extending to approximately $0.1\,\mathrm{s}$} 

Taken together, these results demonstrate that the load balancer effectively handles a workload—with non-trivial inter-task dependencies—spanning five orders of magnitude in task duration without any prior knowledge of task runtimes, thus resolving issues with the existing load balancer implementation.


\section{Conclusion} \label{sec:conclusion}



We have presented an improved SLURM-integrated load balancer for the 
\acrshort{umbridge} framework, targeting the heterogeneous and loosely 
dependent workloads of \acrshort{uq} workflows on \acrshort{hpc}
systems. Tested on a Bayesian inverse problem for the 2011 T\=ohoku earthquake 
source location using \acrshort{mlda} with a hierarchy spanning five orders of 
magnitude in runtime, the load balancer achieves average idle times of 
$\mathcal{O}(10^{-3})\,\mathrm{s}$ across five parallel chains without prior 
knowledge of task runtimes or dependencies, substantially improving on existing 
\acrshort{umbridge} load balancers \cite{11018268}. The recovered 
posterior agrees with the known reference solution
\cite{10.1093/gji/ggu203}, validating the statistical and computational
parts of the workflow.

Future work includes incorporating node utilisation awareness for a more dynamic 
\replaced[id=C]{request dispatch mechanism}{resource allocation}, adding a checkpointing \added[id=C]{and restart} mechanism for resilience in lengthy workflows, and \replaced[id=C]{improving the scalability of the load balancer on workloads that require significantly more parallel servers.}{evaluating the load balancer on gradient-based \acrshort{mcmc} methods that place additional heterogeneous demands on the scheduler.}


\begin{credits}
\subsubsection{\ackname} This work has made use of the Hamilton HPC Service and the COSMA supercomputer of Durham University. The COSMA supercomputer is a DiRAC@Durham facility managed by the Institute for Computational Cosmology on behalf of the STFC DiRAC HPC Facility (www.dirac.ac.uk). The equipment was funded by BEIS capital funding via STFC capital grants ST/K00042X/1, ST/P002293/1, ST/R002371/1 and ST/S002502/1, Durham University and STFC operations grant ST/R000832/1. DiRAC is part of the National e-Infrastructure. We also thank the anonymous reviewers for their comments to improve this work.

\subsubsection{\discintname}
The authors have no competing interests to declare that are relevant to the content of this article.
\end{credits}

%
%
%

\bibliographystyle{splncs04}
\bibliography{CP113_references}

\end{document}